\documentclass[conference]{IEEEtran}
\usepackage{multirow}
\usepackage{blindtext, graphicx}
\usepackage{url}
\usepackage{listings}
\usepackage{graphicx}
\usepackage[font=small]{caption}
\usepackage[utf8]{inputenc}

\title{ScenEval: A Benchmark for Scenario-Based Evaluation of Code Generation}

\author{
\IEEEauthorblockN{Debalina Ghosh Paul, Hong Zhu and Ian Bayley}
\IEEEauthorblockA{School of Engineering, Computing and Mathematics, Oxford Brookes University\\ Oxford OX33 1HX, UK. Email: hzhu@brookes.ac.uk}
}

\begin{document}

\maketitle
\thispagestyle{empty}
\pagestyle{empty}

%%%%%%%%%%%%%%%%%%%%%%%%%%%%%%%%%%%%%%%%%%%%%%%%%%%%%%%%%%%%%%%%%%%%%%%%%%%%%%
\begin{abstract}

In the scenario-based evaluation of machine learning models, a key problem is how to construct test datasets that represent various scenarios. The methodology proposed in this paper is to construct a benchmark and attach metadata to each test case. Then a test system can be constructed with test morphisms that filter the test cases based on metadata to form a dataset.

The paper demonstrates this methodology with large language models for code generation.  A benchmark called ScenEval is constructed from problems in textbooks, an online tutorial website and Stack Overflow. Filtering by scenario is demonstrated and the test sets are used to evaluate ChatGPT for Java code generation.

Our experiments found that the performance of ChatGPT decreases with the complexity of the coding task. It is weakest for advanced topics like multi-threading, data structure algorithms and recursive methods. The Java code generated by ChatGPT tends to be much shorter than reference solution in terms of number of lines, while it is more likely to be more complex in both cyclomatic and cognitive complexity metrics, if the generated code is correct. However, the generated code is more likely to be less complex than the reference solution if the code is incorrect. 
\end{abstract}

\begin{IEEEkeywords}
Machine learning; Large language models; ChatGPT; Code Generation; Benchmark; Performance evaluation; Scenario-based testing.
\end{IEEEkeywords}

%%%%%%%%%%%%%%%%%%%%%%%%%%%%%%%%%%%%%%%%%%%%%%%%%%%%%%%%%%%%%%%%%%%%%%%%%%%%%%%%
\section{Introduction}

Scenario-based testing has long been proven to be an efficient and effective testing method for traditional software and widely applied in practice. For machine learning (ML) applications, standards for developing safety critical systems, like ISO26262 \cite{z3} for road vehicles,  requires the method of scenario-based testing to be applied systematically to thoroughly cover hazardous scenarios. Consequently, recent years have seen a rapid growth in research on scenario-based testing of autonomous vehicles \cite{z1,z2}. 

It is highly desirable, however, that scenario-based testing can be applied not only to conventional software and safety critical applications, but also to sophisticated ML models such as large language models (LLMs)\cite{c1}. A key problem is how to construct datasets that represent various scenarios efficiently and effectively. This paper addresses this problem in the context of testing and the evaluation of an LLM's capability for code generation. 

This paper is organised as follows. Section \ref{secII} reviews existing work on the testing and evaluation of code generation, including benchmarks and the performance metrics. Section \ref{secIII} explains how benchmark ScenEval was constructed and analyses its main characteristics. Section \ref{secIV} presents the datamorphic test system for scenario-based testing with ScenEval and its implementation using the tool Morphy. Section \ref{secV} reports a case study with the testing and evaluation of ChatGPT. Section \ref{secVI} concludes the paper with a discussion of future work.

\section{Related Work}\label{secII}

\subsection{Scenario-Based Testing and Evaluation of ML}

A scenario is an operational situation in a given use case of a system. For traditional software, a scenario is typically represented as a linear sequence of interactions between the user and the system. The identification and specification of scenarios is an integral part of use case driven software engineering \cite{z5}. Test data can then be derived from the sequence of human-computer interactions through instantiation. In contrast, for ML applications, it is often a category of input queries given to the ML model that represents the same operation situation. Therefore, traditional scenario-based testing techniques cannot be applied straightforwardly. 

In order to address this problem,  Zhu \emph{et al}. \cite{z4} proposed a process model for identifying scenarios in the operation of ML applications, and defined a set of test adequacy criteria to cover combinations of scenarios. In \cite{c1}, Zhu \emph{et al}. advanced an automated technique for generating test data. It employs data augmentation operators known as datamorphisms to transform test data of a \emph{seed scenario} to a \emph{mutant scenario}. 

This technique was then applied to deep neural networks (DNN) for computer vision, specifically the perception system of an autonomous racing car.  By evaluating the system on various scenarios, the worst performing scenarios were identified and the DNN re-trained with additional data for those scenarios and its performance improved.

While these experiments demonstrated the effectiveness and efficiency of the approach, its applicability requires the existence of data for \emph{seed scenarios} and datamorphisms to transform them to mutant scenarios. In the case of autonomous vehicles, these are difficult to obtain. Much research efforts have been spent on simulation of different traffic and road conditions \cite{z1,z2}. However, as far as we know, there is little work on constructing suitable benchmarks for scenario-based testing in other ML application domains.

\subsection{Benchmarks for Code Generation}

This paper concerns benchmarks for the evaluation of ML models as code generation tools. Each element of the dataset contains a natural language input that specifies the programming task. The ML model is expected to generate a piece of program code that meets the specification. 

We have identified 11 different such benchmarks in the literature. They differ from each other in the way that the data are procured, the contents contained in each element, the type of code to be generated and the target language. Their key features are summarised in Table \ref{tabI}. 

\begin{table*}[h]
\caption{Main Features of Existing Benchmarks} \label{tabI}
\begin{center}
\begin{scriptsize}
\begin{tabular}{|l|l|l|c|l|c|l|c|c|c|c|}
\hline
\textbf{Benchmark} & \textbf{Source} &  \textbf{Level}& \#\textbf{Tasks} &\textbf{Language} &\textbf{Signature} &\textbf{\#Tests} &\textbf{Solution} &\textbf{Difficulty Levels}\\ \hline
APPS\cite{c8} & Coding challenge &Program  &10,000 &Python &-  &+ &+ &3 \\ \hline
HumanEval\cite{c9}  & Domain Experts &Function &164 &Python &+  &7.7 &- &- \\ \hline
MBPP\cite{c10} & Crowd-sourcing &Function & 974 &Python &-  &3 &- &- \\ \hline
MathQA-Python\cite{c10} & MathQA &Function &23,914 &Python &-  &3 &- &- \\ \hline
ClassEval\cite{c14}  & Repository, HumanEval, MBPP &Class &100 &Python &+  &33.1 &+&- \\ \hline
CoderEval\cite{c5}  & Github &Function & 230 &Python &+  &+ &-&6 \\ 
                  & &Method &230 &Java & & &&\\ \hline
Multipl-E\cite{c11}  & HumanEval, MBPP  &  Function & 1138 &Various &+  &3 to 7 &-&- \\ \hline
DS-1000 \cite{c7}  & Stack Overflow &Statement &1000 &Python &+  &1.6 &+&-\\ \hline
HumanEval+\cite{c12}  & HumanEval &Function & 164 &Python &+ &774.8 &-&- \\ \hline
CONCODE \cite{c6}  & Github &  Method & 2000 &Java &+  &- &-&- \\ \hline
R-benchmark\cite{c15}  & Text Books & Program &351 &R &-  &- &+&3 \\ \hline
\end{tabular}
\end{scriptsize}
\end{center}
\end{table*}

The benchmarks differ in the contents of each element of the dataset. Natural language descriptions are always present, usually given as docstrings. In addition, function signatures and unit test cases may be provided, both of which are used for test automation to check the correctness of generated solutions. In some cases, there may also be reference solutions. Table \ref{tabI} also gives the contents provided by each benchmark; the column \#Tests gives the average number of test cases per task if any are provided. In some cases, programming tasks in the dataset are classified into subsets of different difficulty levels. The column Difficulty levels shows the number of difficulty levels that the test cases were classified into.  

%For example, Hendrycks et al. \cite{c8} distinguish three levels (Introductory, Interview, Competition) in APSS.  Similarly, Austin et al. \cite {c10} has two levels associated with  subsets MBPP and MathQA-Python. Similarly, Yu et al. \cite{c5} used, for CoderEval, six levels according to the function's contextual dependency. 
%Miah and Zhu \cite{c15} also distinguish three difficulty levels but as a part of the metadata associated to each task. 

\subsection{Evaluation of Code Generation Capability}

In general, the evaluation of a ML model involves activities at two levels: the individual test case level, where the ML model's output on each test case is evaluated, and the benchmark dataset level, where the overall performance is calculated from the assessments of the individual test cases.
 
%In our case, the ML model is a code generator. Thus, the test cases are coding tasks and the outputs are program code, which could be accompanied by text explanations sometimes. Various techniques and methods have been developed to assess the quality of the output from a LLM on a coding task and to calculate the overall performance on a benchmark. 

There are two different approaches for evaluating the quality of code. The first approach is to measure its syntactic closeness to a reference solution,  which can be done with the BLEU metric.
%Correctness has been the main quality attribute of assessing the response of the LLM to each individual programming task. According to what is provided by the benchmark, this is done either with reference solutions (ConCode), or using test cases (HumanEval, HumanEval+, MBPP, MathQA-Python and MultiPL-E) or a combination of both (APPS, ClassEval, CoderEval, and DS-1000). 
However, Kulal et al. \cite{c17} found that BLEU fails to reflect  functional correctness and Hendrycks et al. \cite{c8} showed that it is even inversely correlated with it. In 2020, Ren et al. \cite{c18} introduced, as an alternative to BLEU, the measure CodeBLEU, which compares abstract syntax trees and data flow graphs instead of program text. Lai et al. used a much-relaxed form of similarity metric called surface-form constraints, which identifies keywords and the presence or absence of certain APIs \cite{c7}. 

The second approach, proposed by Kulal et al., is to measure functional correctness instead and regard generated code as correct if it passes all test cases; Hendrycks et al.  \cite{c8} , in contrast, measured the \textit{percentage} of test cases passed.

Kulal et al. measured the overall performance with the percentage of coding tasks for which the LLM produces at least one correct solution when asked to produce 100 solutions. This is later generalised to  $k>0$ solutions for each task and the definition of the $pass@k$ metric, which is the probability of generating at least one solution in $k$ successfully. Chen et al. \cite{c9} found that the $pass@k$ metrics produce a high variance, however, so they counted the number $c$ of successful solutions in $k$ and used $c$ and $k$ to make an unbiased estimation of the $pass@k$ metric.  That has been used by most of the benchmarks reported above. 

Miah and Zhu \cite{c15}, in contrast, considered the use of a LLM model to be an interactive process in which the user makes a number of attempts by entering and revising the input to the LLM until a successful solution is generated, or gives up after a maximal number $k$ of allowed attempts. This is notably different from $k$ different solutions of one attempt. They proposed a new metric $\#attempt_k$, which is the average number of attempts over the benchmark. 

Table \ref{tabV} provides detailed information regarding the benchmarks used in the evaluation, the metrics used and the main results. 

\begin{table}[h]
\caption{Uses of Benchmark In Evaluations}
\label{tabV}
\begin{center}
\begin{scriptsize}
\begin{tabular}{|l|l|l|l|c|}
\hline
\textbf{Benchmark} &\textbf{Correctness} &\textbf{Perf. Metrics} &\textbf{ML Model} &\textbf{Result}\\
\hline
APPS &  pass all tests,    & \%pass@100,  &GPT-2 &0.68, 7.96\\
           &  \%passed tests &Avg \%passed  & GPT-Neo &1.12, 10.15\\
           &                           &                          &GPT-3 &0.06, 0.55  \\
\hline
HumanEval & pass all tests & \%pass@100 &GPT-Neo &21.37\\
                   &                       &                    &Codex &72.31\\
\hline
MBPP  & pass all tests & \%pass@1 &Decoder  &79.0 \\
			  &                      &                    &Transformer &82.8 \\
			  &                      &                    &Lang. Model &83.8 \\
\hline
MathQA & pass all tests & \%pass@1 &Decoder &74.7\\
-Python &                        &                   &Transformer &79.5 \\
             &                         &                   &Lang. Model  &81.2 \\
\hline
ClassEval & pass all tests & \%pass@5 &  GPT-4 & 42.0 \\
                 &                       &                   &  GPT-3.5 &36.0 \\
                 &                       &                   &  CodeGen &13.0 \\
\hline
CoderEval & pass all tests & \%pass@10 &CodeGen &23.48 \\
(Python)    &                       &                      &ChatGPT &30.00 \\  
\hline
CoderEval &pass all tests & \%pass@10 &CodeGen &33.48 \\  
(Java)        &                       &                     &ChatGPT &46.09 \\  
\hline
Multipl-E        &pass all tests & \%pass@1 &Codex &$\approx$ 36 \\
(HumanEval) &                      &                   &CodeGen &$\approx$ 9\\
\hline
Multipl-E  & pass all tests & \%pass@1 &Codex & $\approx$40\\
(MBPP)    &                      &                    &CodeGen & $\approx$14 \\  
\hline
DS-1000 & pass all tests & \%pass@1 &Codex-002 &41.25 \\  
               &                       &                    &CodeGen &8.4 \\  
\hline
HumanEval+ & pass all tests & \%pass@100 &CodeGen &$\approx$64.0 \\  
                      &                      &                        &ChatGPT & 89.8\\  
                      &                      &                        &GPT-Neo &16.8 \\
\hline
ConCode & BLEU & Avg BLEU &Retrieval &20.27 \\  
                 &            &                  &Seq2Seq &23.51 \\  
                 &            &                  &Seq2Prod &21.29 \\  
\hline
R-benchmark & Satisfactory & Avg \#attempt$_k$ & ChatGPT & 1.6\\
\hline
\end{tabular}
\end{scriptsize}
\end{center}
\end{table}

The only quality attribute considered above is correctness except that a work by Miah and Zhu \cite{c15} assessed structuredness, conciseness, completeness, and logical clarity,  as well as attributes on the textual explanation of the code.

In summary, existing benchmarks for code generation fall short in their support for scenario-based testing since none of them link test cases to scenarios that might be encountered during code generation. The solution proposed by Miah and Zhu \cite{c15} was to include metadata with each test case to represent the scenario it tests. 

This paper further develops that metadata approach by constructing a large scale benchmark ScenEval with 12864 Java programming tasks, of many different kinds and from different sources, all tagged with scenario information as metadata in JSON format. With the support of the automated testing tool Morphy \cite{c2}, the concept of test morphisms makes it easy to form datasets with test cases that all belong to one scenario and, by changing and combining datasets, to study the performance of a LLM across several different scenarios.

\section{ScenEval Benchmark}\label{secIII}

We now describe the structure of each task within ScenEval and then describe the scenarios it handles. All tasks are labelled with scenario information to improve upon existing work and for the same reason, a variety of sources have been used.

\subsection{Structure of Data}

\begin{figure}
\centering
\includegraphics[width=8.5cm]{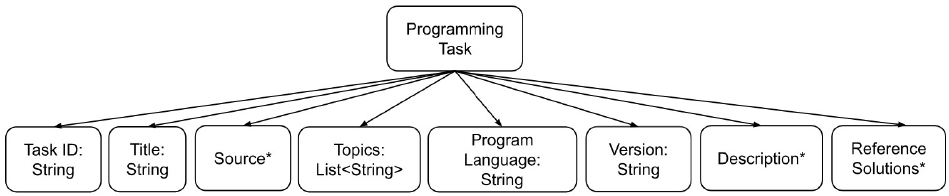}
\captionsetup{justification=centering, position=above}
\caption{Structure of JSON representation of Task}
\label{fig1}
\end{figure}

\begin{figure}
\centering
\includegraphics[width=8.5cm]{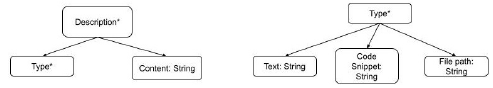}
\captionsetup{justification=centering, position=above}
\caption{Structure of JSON representation of Description and Type}
\label{fig4}
\end{figure}

In ScenEval, each test case is a coding task with the following metadata represented as a JSON value whose structure is given in Figure \ref{fig1}. 
\begin{itemize}
\item Task Id: an universal unique identifier of coding task;
\item Title: the title of the coding task;  
\item Source: a list of sources, which can be more than one if the task occurs in more than one sources. 
\item Topics: a list of topics covered by the coding task.
\item Programming language: The programming language in which the code is to be generated. 
\item Version: The version number of the task, to support the  evolution of the dataset. 
\item Description: A description of the coding task, which can be a sequence of text or a code snippet (such as a signature or a skeleton), or a fully qualified file name for image data; see Figure \ref{fig4}. 
\item Reference Solutions: A list of reference solutions. 
\end{itemize}

\begin{figure}
\centering
\includegraphics[width=8.5cm]{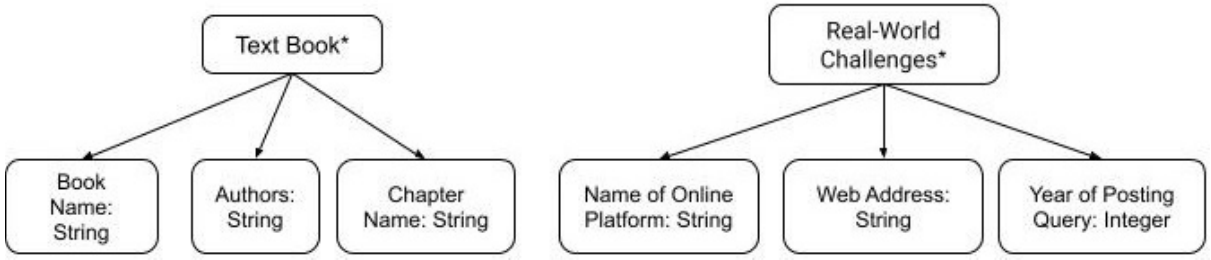}\\
\includegraphics[width=8.5cm]{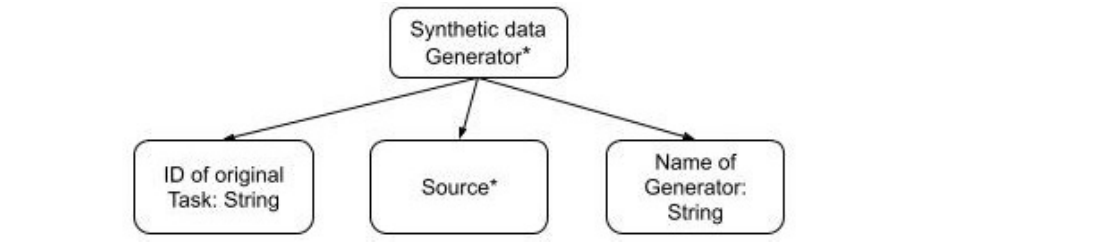}
\captionsetup{justification=centering, position=above}
\caption{Structure of JSON Representation of Various Types of Sources}
\label{fig3}
\end{figure}

Currently, we distinguish three types of sources for the coding tasks: textbook, real-world questions, and synthetic data.  The metadata structure for each source type is given in Figure \ref{fig3}. More types of sources can be easily added due to the extensibility of JSON. 

%\begin{figure}
%\centering
%\includegraphics[width=8.5cm]{fig2.jpg}
%\captionsetup{justification=centering, position=above}
%\caption{Structure of JSON representation of Source}
%\label{fig2}
%\end{figure}

%\footnote{To assess the semantic consistency of the generated code and robustness of language models, we intend to integrate synthetic data generators such as QuillBot.}

%The representation of task description can be seen in Figure \ref{fig4}. The content attribute can include diagrams in the form of the fully qualified file name under which it is saved.

\begin{figure}[h]
\centering
\includegraphics[width=6cm]{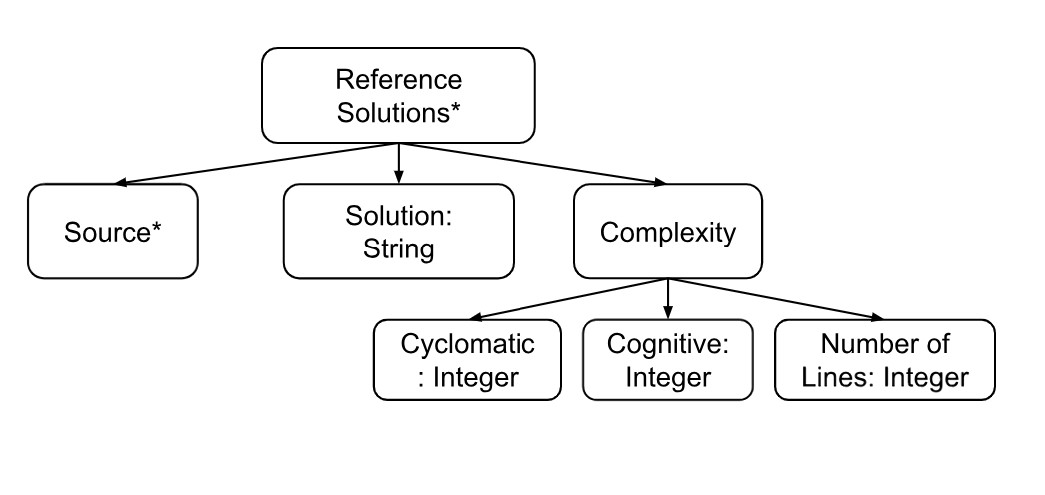}
\captionsetup{justification=centering, position=above}
\caption{Structure of JSON Representation of Reference Solutions}
\label{fig5}
\end{figure}

We allow multiple reference solutions to be provided for each coding task. As shown in Figure \ref{fig5}, each reference solution is also associated with metadata for its source and complexity. Three metrics are used for the latter: cyclomatic complexity, cognitive complexity and the number of lines.

%Cyclomatic complexity is the number of linearly independent paths in the code. Cognitive complexity measures how difficult it is for humans to comprehend the program. For example, nested control flows will make that higher. Both types of complexity are positively correlated with understandability [45????] as high cyclomatic complexity indicates excessive branching. Both are measured with the PMD code analyzer\footnote{URL: https://pmd.github.io/}. The number of lines is also included as a measure of conciseness.

\begin{figure}
\centering
\includegraphics[width=8.5cm]{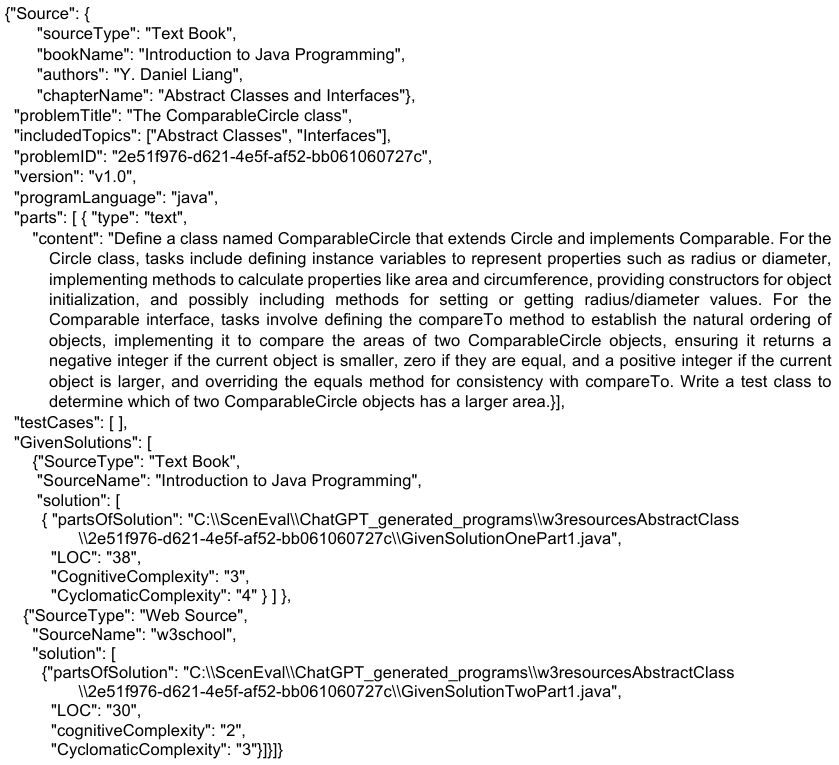}
\captionsetup{justification=centering, position=above}
\caption{Example of a Test Case}
\label{fig32}
\end{figure}

Figure \ref{fig32} shows  an example of the test cases in ScenEval.  

\subsection{Data Procurement and Extraction}

The tasks of ScenEval are extracted from three types of sources:
\begin{itemize}
    \item \emph{Textbook}: Exercises and solutions in four textbooks on Java programming \cite{c19}, \cite{c20}, \cite{c21}, \cite{c22}. 
    \item \emph{Online learning website}: Exercise questions and solutions on Java programming at the website of W3Resources\footnote{URL: https://www.w3resource.com/java-exercises/}.
    \item \emph{Online forum}: Questions and answers about Java programming posted on Stack Overflow\footnote{URL: https://stackoverflow.com/}.  
\end{itemize}

Data in the textbooks are extracted manually and metadata are also assigned manually. Exercise questions that do not require code to be written are excluded. There are a total of 1306 tasks from 4 textbooks; see Table \ref{tabVI} for the number of tasks extracted from each source. 

\begin{table}[h]
\caption{Number of Tasks from Various Sources}
\label{tabVI}
\begin{center}
\begin{tabular}{|p{7cm}|c|}
\hline
\textbf{Source} & \textbf{\#Tasks} \\ \hline
W3Resources &1058\\ \hline
Stack Overflow &10500 \\ \hline
Textbooks & 1306\\ \hline
\scriptsize{Introduction to Java Programming by Y. Daniel Liang} &\scriptsize{791}\\
\scriptsize{Absolute Java by Walter Savitch} & \scriptsize{217}\\
\scriptsize{Java: A Beginner’s Guide by Herbert Schildt} & \scriptsize{230}\\
\scriptsize{Programming and Problem Solving with Java by Nell Dale et al.} &\scriptsize{68}\\ \hline
\textbf{Total} & \textbf{12864} \\ \hline
\end{tabular}
\end{center}
\end{table}

For the online sources, data together with the metadata are extracted automatically by running script codes. Tasks from W3Resource have questions that are well presented as exercises for students who are learning programming and the solutions are tested and reliable. Therefore, in the sequel, they are also categorised as \emph{textbook questions}. 
%Figure \ref{fig6} shows the distribution of topics within this category. 

%\begin{figure}[h]
%\centering
%\includegraphics[width=8.5cm]{figs/fig13b.pdf}
%\captionsetup{justification=centering, position=above}
%\caption{Distribution of Textbook Questions by Topic}
%\label{fig6}
%\end{figure}

The questions from Stack Overflow are also extracted automatically, but manually inspected to remove those not on code generation, but apart from that, neither the questions nor their solutions were edited, since we believe it is best to keep them both in their natural form for testing LLMs, even though the solutions are untested and there is little quality control other than user votes and feedback. These tasks are referred to as \emph{real-world questions} in the sequel. 

\begin{figure}[h]
\centering
\includegraphics[width=8.5cm]{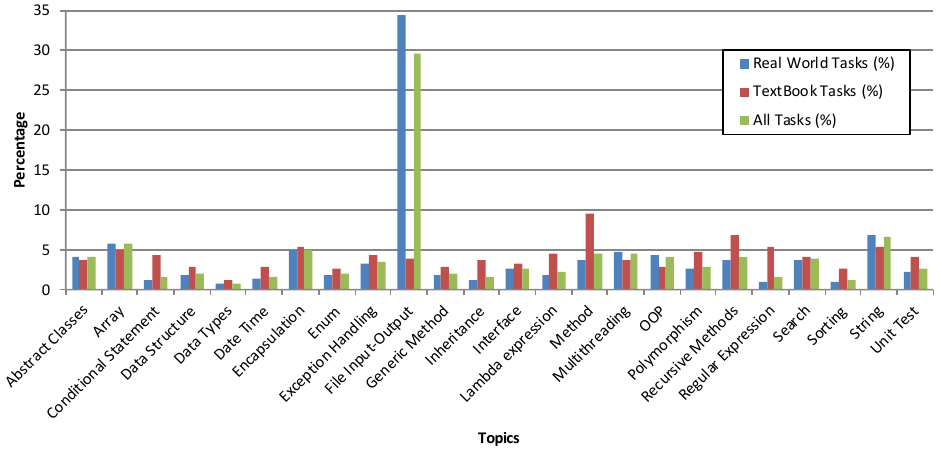}
\captionsetup{justification=centering, position=above}
\caption{Distribution of Coding Tasks by Topic}
\label{fig7}
\end{figure}

\begin{figure}[h]
\centering
\includegraphics[width=7.5cm]{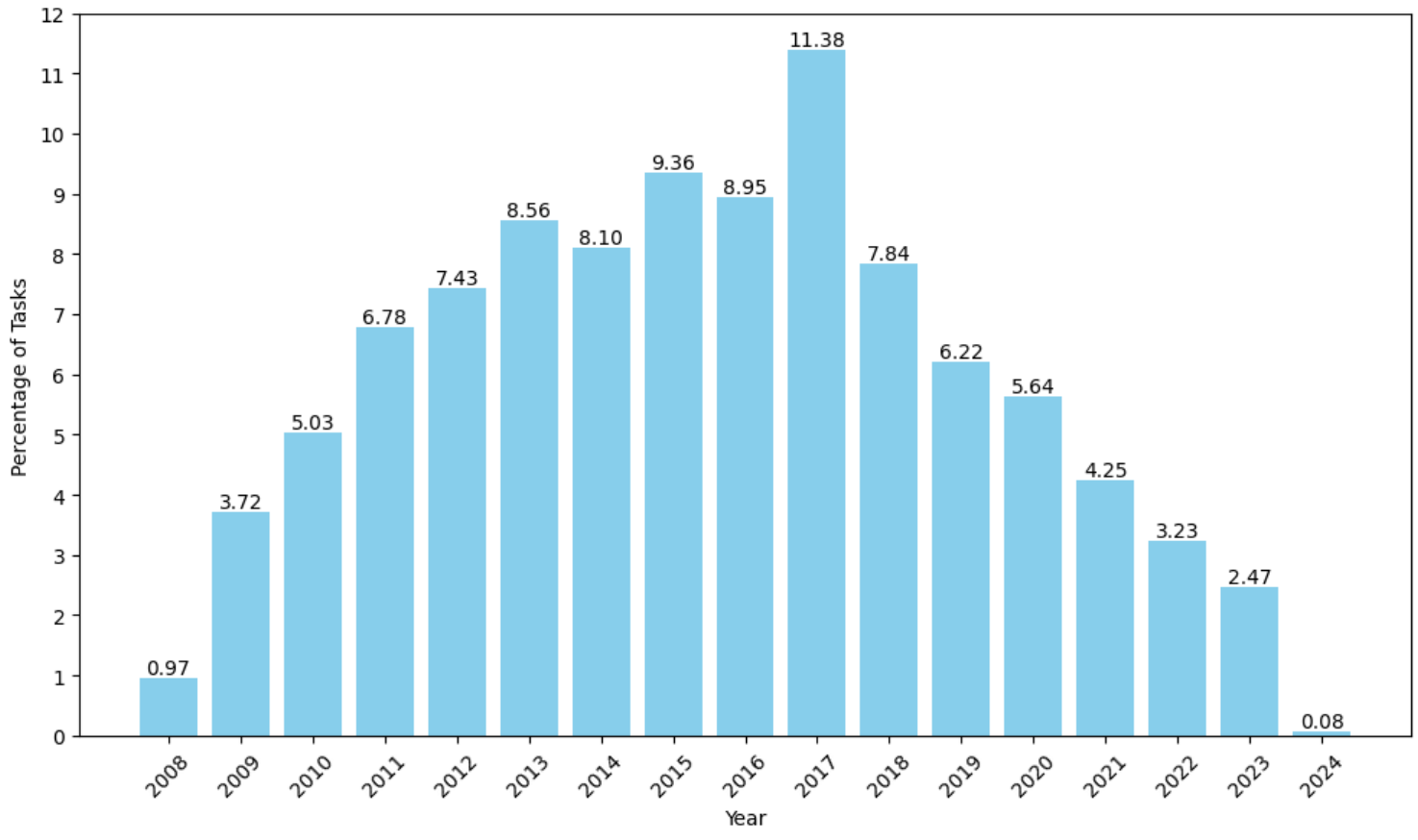}
\captionsetup{justification=centering, position=above}
\caption{Distribution of Real-World Questions by Year}
\label{fig8}
\end{figure}

Figure \ref{fig7} shows the distribution of coding tasks across different topics. It is worth noting that the real-world questions were posted to Stack Overflow over a period of time since 2008. Figure  \ref{fig8}  shows the distribution of these tasks across the time. 

%The Stack Overflow tasks are professional scenarios and since the metadata includes topics and year of posting, we can filter either according to year (as shown in figure 7) or both year and topic (as shown in figure 8) . 

%In contrast, the tasks sourced from textbooks represent academic scenarios. We can filter those using metadata such as topics and complexity, as demonstrated in figures 11, 12 and 13.

\begin{figure}[h]
\centering
\includegraphics[width=8.5cm]{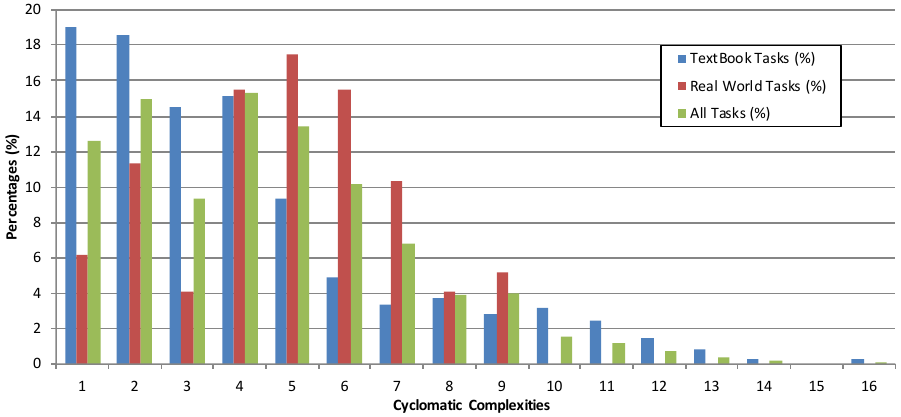}\\
\includegraphics[width=8.5cm]{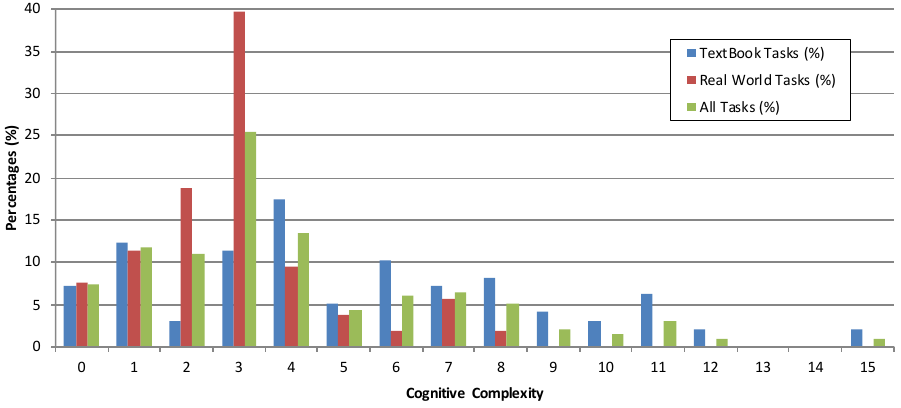}\\
\includegraphics[width=8.5cm]{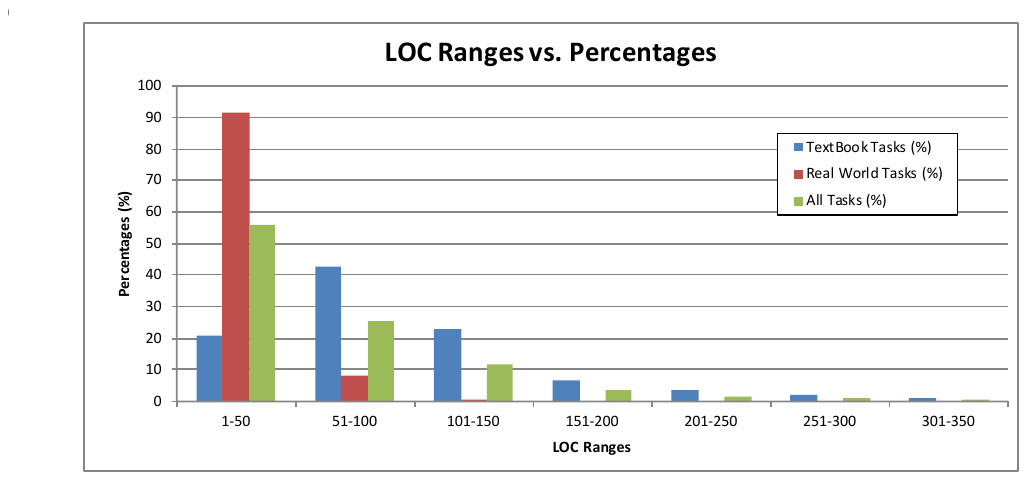}
\captionsetup{justification=centering, position=above}
\caption{Distribution of Coding Tasks}
\label{fig14b}
\end{figure}

Figure \ref{fig14b} shows the distributions of tasks by complexity according to cyclomatic complexity, cognitive complexity and lines of code (LOC), respectively. 

\section{Datamorphic Test System}\label{secIV}

The methodology of datamorphic testing regards software testing as a problem of systems engineering, and aims to develop and apply a test system, in which testing is performed, test resources are managed and testing processes are automated.

Moreover, datamorphic testing constructs test systems by classifying the artefacts of testing into two types: test entities (such as test data, objects, the software under testing, test results, etc.) and test morphisms, which are operations on or transformers of test entities. Here, benchmarks are test entities and so are the various subsets of a benchmark that represent various scenarios.

The tool Morphy has been developed to support datamorphic testing \cite{c2}. The tester can  define test entities in Java as classes and implement test morphisms as Java methods. In addition to general test actions, Morphy recognises the following types of test morphisms:
\begin{itemize}
\item \emph{Seed maker}: which generates test cases from other types of entities;
\item \emph{Datamorphism}: which transforms test cases;
\item \emph{Metamorphism}: which checks the correctness of test cases and output a Boolean value; 
\item \emph{Test set filter}: which add or remove test cases from a test set;
\item \emph{Test set metric}: which maps a test set to a real value, such as its test adequacy;
\item \emph{Test case filter}: which maps a test case to a Boolean value to decide whether to keep it in the test set;
\item \emph{Test case metric}: which maps test cases to a real value, such as its complexity. 
\item \emph{Analyser}: which analyses the test set and produces a test report.
\item \emph{Executer}: which invokes the program under test with input data from the test case and receives the output from the program. 
\end{itemize}

In addition to providing facilities to manage test entities, Morphy supports test automation at three levels, each of which can be invoked with a button click:
\begin{itemize}
    \item  \emph{action}, the level of a single test activity;
    \item \emph{strategy}, a composition of test morphisms, which can be defined as an algorithm that contains test morphisms as parameters;
    \item  \emph{process}, a sequence of invocations of actions and strategies that can be recorded into an editable test script. 
\end{itemize}

To support scenario-based testing in particular, we defined and implemented four different sorts of test morphisms: test set filters, analysers, seed makers and test executors. We will now examine each of these.

%mappings on test entities, which can include mappings such as filters from one set of entities (the whole benchmark) to another set (the tasks belonging to the desired scenario). Other morphisms execute tests to capture results and analyse results to produce test reports, which we discuss in the next section.  The testing process can be scripted so that manual intervention is not needed.

\subsection{Test Set Filters.}

Four test set filters have been implemented to select test tasks in a dataset according to the source, topics, complexity and years. 

\begin{itemize}
\item \emph{filterBySource}, which filters test cases according to the source type supplied as input; if that source type is 'text book', the user can select a specific textbook

\item \emph{filterByTopic}, which filters test cases according to a set of topics supplied as input

\item \emph{filterByComplexity}, which filters test case according to complexity metric and a range within that metric 

\item  \emph{filterByYears}, which filters test cases according to a range of years, specified with a start year and end year.
\end{itemize}

These filters can be combined. For example, you can create a dataset with topics on threads from real-world questions after the year 2010, by applying filterByTopic then filterBySource then filterByYear.  Datasets can be saved and then loaded to be combined with other datasets.

\subsection{Test Data Analysers.}

There are two types of analysers, those that analyse the data distributions and those that analyse each test case on various quality attributes.

\subsubsection{Analysers of data distributions.} 

Three analysers have been implemented to calculate and display the data distribution in the dataset according to the topic, year and complexity respectively: 

\begin{itemize}
\item \emph{TopicBasedDistribution}, example outputs of which can be seen in Figure \ref{fig7}.

\item \emph{YearBasedDistribution}, as seen in Figure \ref{fig8}. 

\item \emph{ComplexityBasedDistribution}, as seen in Figure \ref{fig14b}. 
\end{itemize}

\subsubsection{Analysers of test cases features.}

\begin{figure}[h]
\centering
\includegraphics[width=7cm]{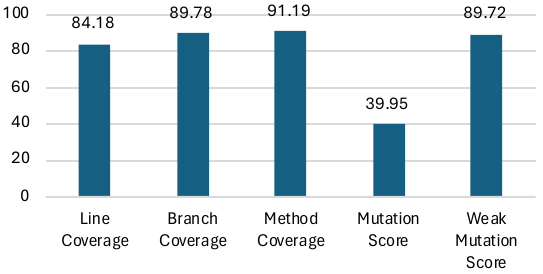}
\captionsetup{justification=centering, position=above}
\caption{Average Test Coverages}
\label{fig10}
\end{figure}

Seven analysers have been implemented to analyse various aspects of the quality of each test task in the dataset. 
\begin{itemize}
\item \emph{isCodeCompilable}, which takes the generated code snippet, compiles it and returns a message saying whether the compilation was successful.

\item \emph{isCodeExecutable}, which takes the object code and executes it to show the output.

\item \emph{analyseComplexity}, which calculates the cyclomatic complexity and cognitive complexity of the solution and saves it in the metadata along with the number of lines. The function is implemented by invoking the PMD code analyser\footnote{URL: https://pmd.github.io/} and extracting the output produced. 

\item \emph{generateTestCode}, which generates two JUnit test classes, one for the reference solution and the other for the generated solution. It is implemented by invoking Evosuite \cite{c3, c4}. 

\item \emph{purifyReferenceTestCode}, which runs all unit tests on the reference solution, reports the correctness of each one and removes the failing test cases

\item \emph{purifySolutionTestCode}, as for \emph{purifyReferenceTestCode}, but applied to the solution test code. 

\item \emph{runTestCodes}, which runs the test cases on both the reference solutions and generated solutions, reports on the correctness of the latter, and then measures test adequacy according to various code coverage metrics by invoking the functions provided by Evosuite \cite{c3, c4}. Figure \ref{fig10} shows the average test coverage of the reference solutions on the whole ScenEval benchmark as an example of the output from this analyser.
\end{itemize}

\subsection{Seed Makers}

Three seed makers have been implemented.  

\begin{itemize}
\item \emph{ManualTaskEntry}, for manually entering test tasks, as used for those in textbooks. 
\item \emph{ExtractStackOverflow}, to extract tasks from Stack Overflow. 
\item \emph{ExtractW3Resource}, to extract data from W3Resource. 
\end{itemize}

Once the data has been extracted and stored in Morphy, it can be inspected and those tasks not suitable for code generation can be removed. Data can be edited, but it has not been for the current version of the benchmark.

\subsection{Test Executer}

Only one test executer has been implemented:  \emph{ChatGPTTestExecuter}, which invokes ChatGPT through its API, sends requests taken from the test task descriptions and extracts data from the responses. Other LLMs can be tested by writing their own executers which will have different APIs, URL and data formats for request and response messages. 

\section{Evaluation of ChatGPT}\label{secV}

This section demonstrates the use of ScenEval in forming various subsets with the support of Morphy so that a scenario-based evaluation can be carried out on ChatGPT.

\subsection{Design of The Experiments}

\subsubsection{Test Case Generation}

Test cases are generated by applying the \emph{generateTestCode} test morphism. From the reference solution,
%by applying the test morphism \emph{generateTestCodes}
it produces a JUnit test class with test cases that come with expected outputs that are checked using assertions. The test cases are not always correct, however, so the ones that failed on the reference solution are removed, by invoking the test morphism \emph{purifyReferenceTestCode}. 

It is worth noting that test case generation from the reference solution only uses the information contained in the reference code. It can be effective to check the correctness of a generated solution on the domain that the reference solution defines. Defects of a generated solution on this domain are called \emph{omission errors} in software testing literature. However, a generated solution may also provide functions beyond this required domain, such as those containing malicious code. The test cases generated from the reference solution cannot detect such defects, which are called \emph{commission errors}. In other words, a test code generated from the reference solution is effective to detect omission errors, but not commission errors. 

To address this problem, we also generate the second test code, but from the generated solution. Similarly, the test morphism \emph{purifySolutionTestCode} is applied to the generated code to remove the incorrect test cases. This test code when applied to the reference solution can detect commission errors. 

In the sequel, the purified the test code generated from the reference solution is called the $\gamma$ test, while the purified test code generated from the generated solution is called $\kappa$ test. 

\subsubsection{Test Execution And Result Analysis}

Once the $\gamma$ and $\kappa$ test codes are ready for a test task, both $\gamma$ and $\kappa$ test codes are applied to both reference solutions and generated solutions. This is by invoking the test morphism \emph{runTestCodes}, which executes the test codes and reports the test results together with the test adequacy measured in various coverage metrics. 

\subsubsection{Correctness Criteria and Performance Metrics}

We use two correctness and performance metrics to measure the performance of ChatGPT's code generation capability. The first is passing all test cases as the correctness criterion and percentage of $pass@1$ as the overall performance metric. Here, a generated solution is regarded as correct by passing all test cases, if both the reference and the generated solutions pass all test cases in the $\gamma$ and $\kappa$ tests. 

The other performance metric is the average pass rate over all tasks, where pass rate for a task is calculated from the failure rate by the formula $1- failure rate$. The failure rate for a task is the proportion of test cases on which either reference solution fails or the generated solution fails over the total number of test cases in $\gamma$ and $\kappa$ tests. 

\subsection{Correctness Analysis}

Our main research goal is to gain insight on how ChatGPT performs on textbooks and real-world questions. Applying the scenario-based evaluation techniques, we created two test datasets; one with tasks from the textbook questions and the other with tasks from the real-world questions by applying the test morphism \emph{filterBySources}. Then, ChatGPT is tested on these two datasets. 
On textbook tasks, the percentage of $pass@1$ is  75.64\%, i.e. about three-quarters of the tasks correctly passed all test cases, with an overall average pass rate of 82.4\%. In contrast, the percentage of $pass@1$ for real-world tasks is 67.07\%, and the overall average pass rate was 74.34\%.

\begin{figure}[h]
\centering
\includegraphics[width=8.5cm]{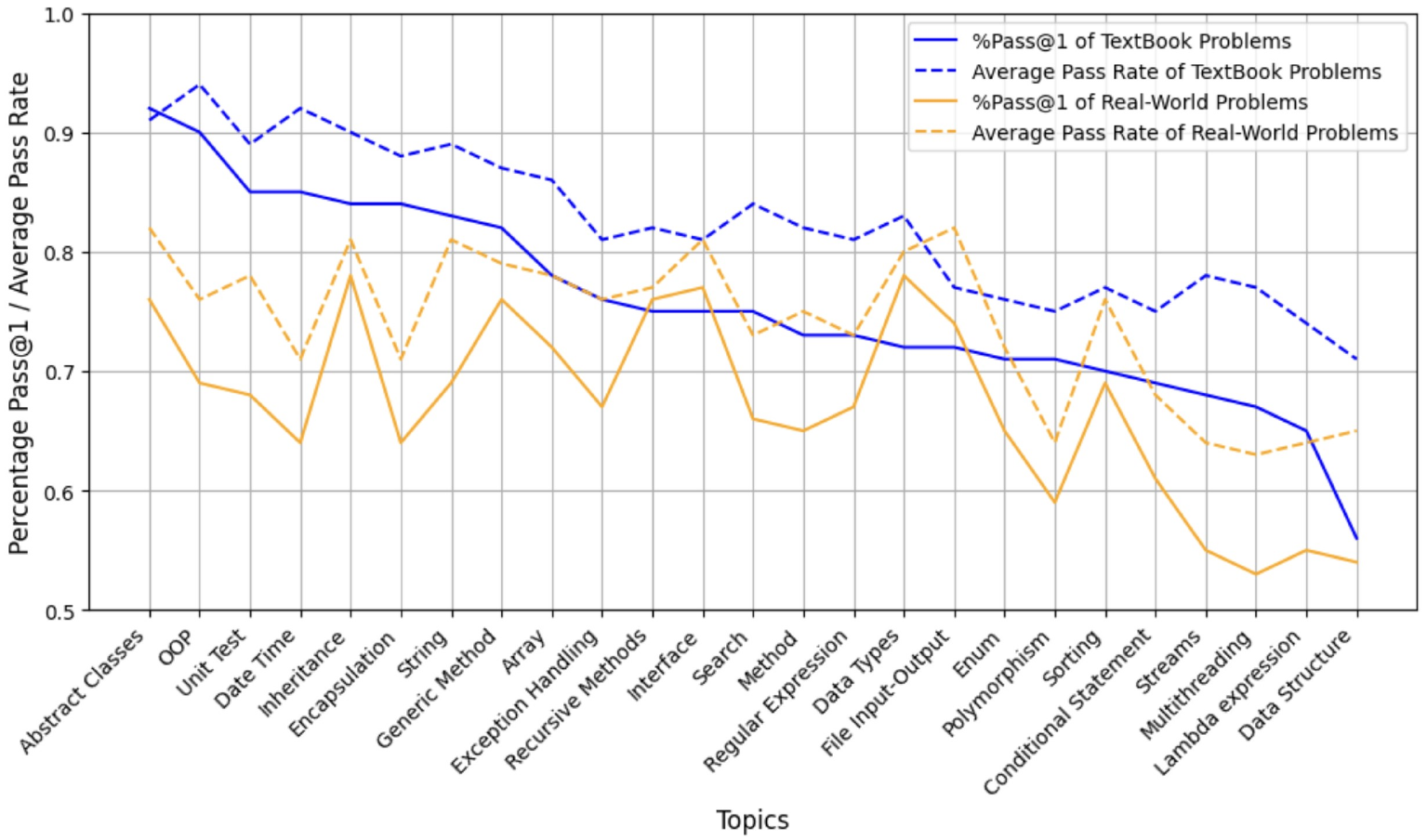}
\captionsetup{justification=centering, position=above}
\caption{Variation of Performance Over Topics}
%Average pass@1 rate and percentage of passed test cases across Academic Scenarios filtered on Topics}
\label{fig18_25}
\end{figure}

\begin{figure}[h]
\centering
\includegraphics[width=8.5cm, height=3.5cm]{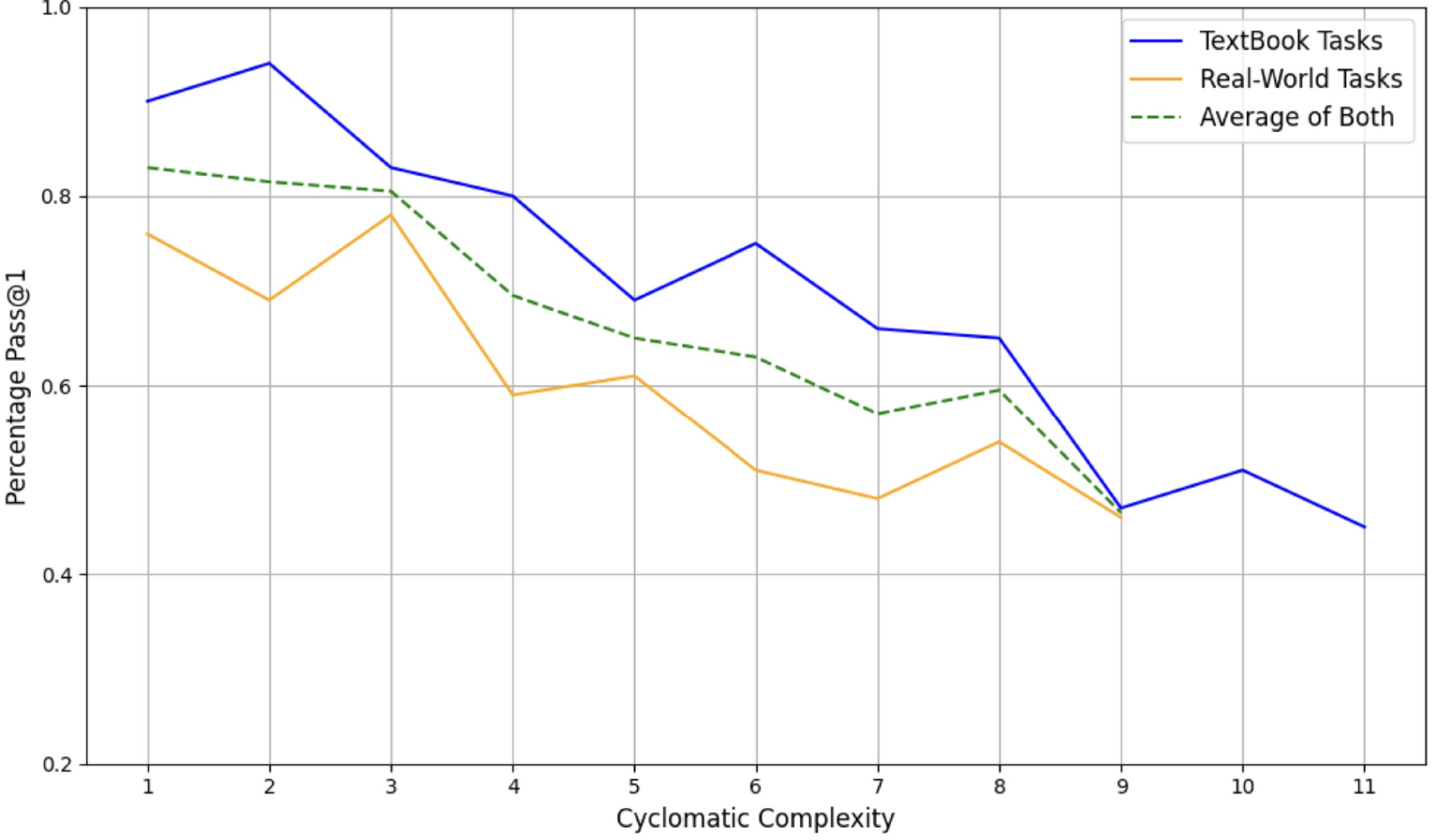}
\captionsetup{justification=centering, position=above}
\caption{Variation of Performance over Cyclomatic Complexity} 
\label{fig21}
\end{figure}

In addition to the overall performances on these two scenarios, we further analysed ChatGPT's performance on tasks of various topics and complexity. Again, by applying the principles of scenario-based evaluation, we split the two test datasets according to topics and complexities using the test morphisms \emph{filterByTopics} and \emph{filterByComplexity} and obtained two sets of sub-datasets. The performances of ChatGPT is evaluated on the datasets. The results are shown in Figure \ref{fig18_25} and \ref{fig21} respectively.  

%The first sub-scenarios we are consider are different topics, as shown in Figure \ref{fig11}. Pass rates vary from a high of 97\% to a low of 56\%. Note that high pass rates correlates with high proportions of passed unit tests. 

%Similar to the analysis on topics, we created two sets of sub-datasets according to the cyclomatic complexity and evaluated ChatGPT on each sub-dataset. The results are shown in Figure \ref{fig21}. 

As shown clearly in Figure \ref{fig21}, ChatGPT's performance decreases as the cyclomatic complexity of the task increases. It might be just a coincidence that some topics are more complicated than others, however, so we pick the worst-performed topics and see whether we observe the same phenomenon.

From the results of the evaluation shown in Figure \ref{fig18_25}, we can identify that ChatGPT performed worst on the 
topics of streams, multi-threading, lambda expressions and data structures. 

\begin{figure}[h]
\centering
\includegraphics[width=8.5cm, height=3cm]{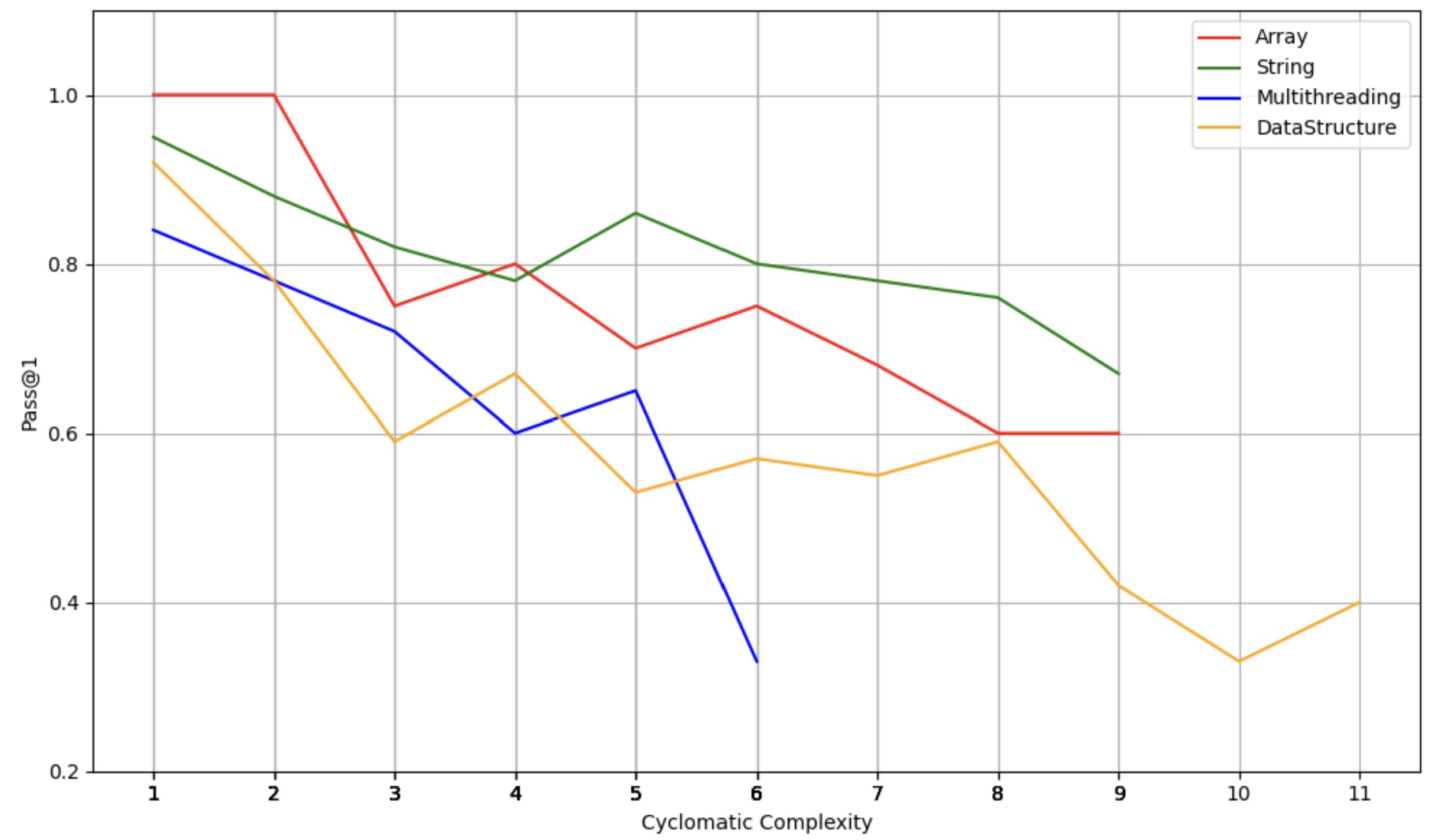}\\
\scriptsize{(a) Performance on Textbook Questions.} \\
\includegraphics[width=8.5cm, height=3cm]{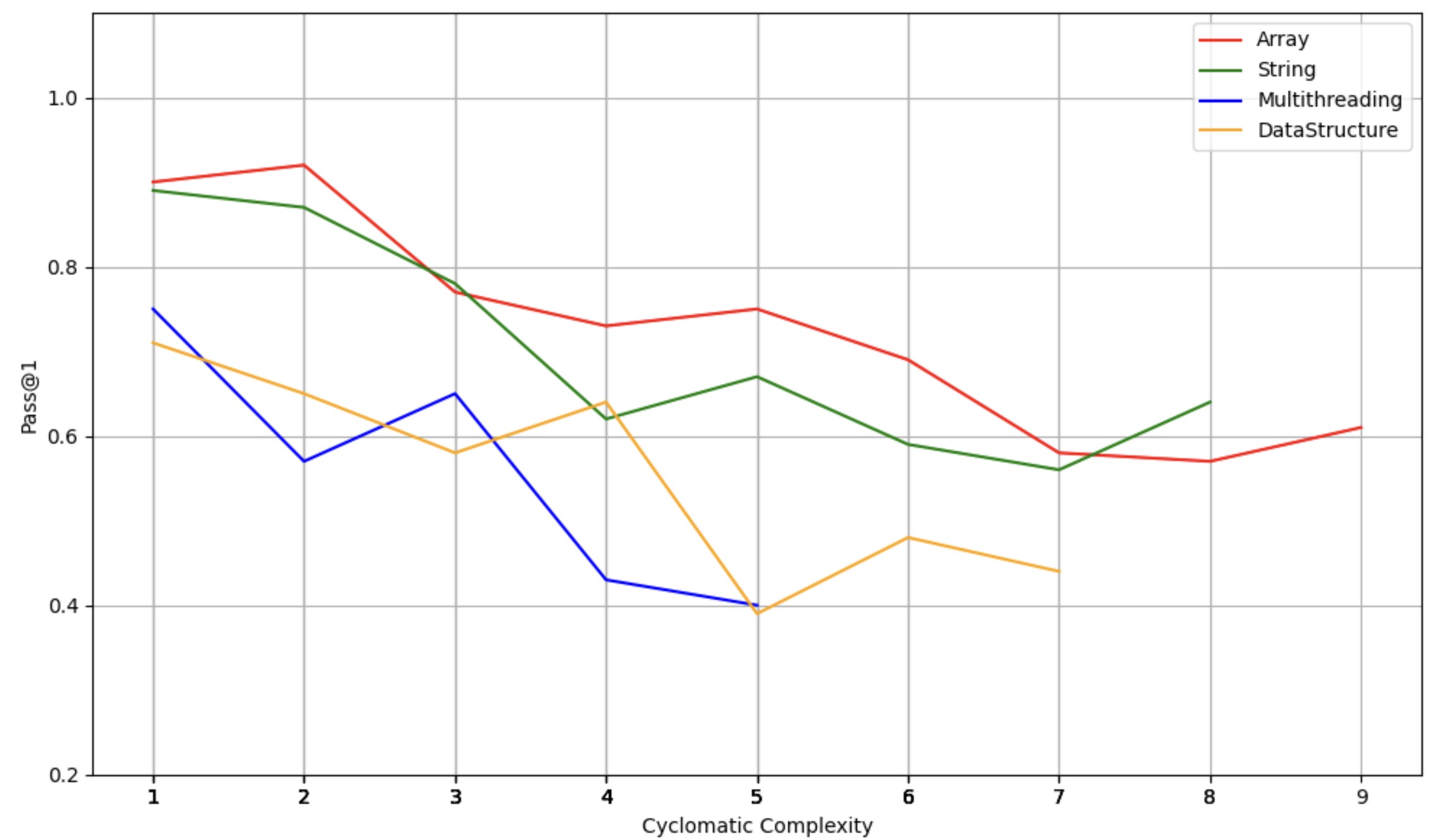}\\
\scriptsize{(b) Performances on Real-World Questions.}\\
\captionsetup{justification=centering, position=above}
\caption{Variation of Performance Over Complexity on Hard Topics}
\label{fig20}
\end{figure}

To gain further insight on ChatGPT's performance on these topics, we split each dataset of these topics into a set of sub-datasets according to the cyclomatic complexity. The evaluation on these sub-datasets revealed that the decline of performance with complexity is uniform for all topics; see Figure \ref{fig20}. Therefore, it is not a coincidence. 

\subsection{Complexity Analysis}

\begin{figure}[h]
\centering
\includegraphics[width=8.5cm, height=3cm]{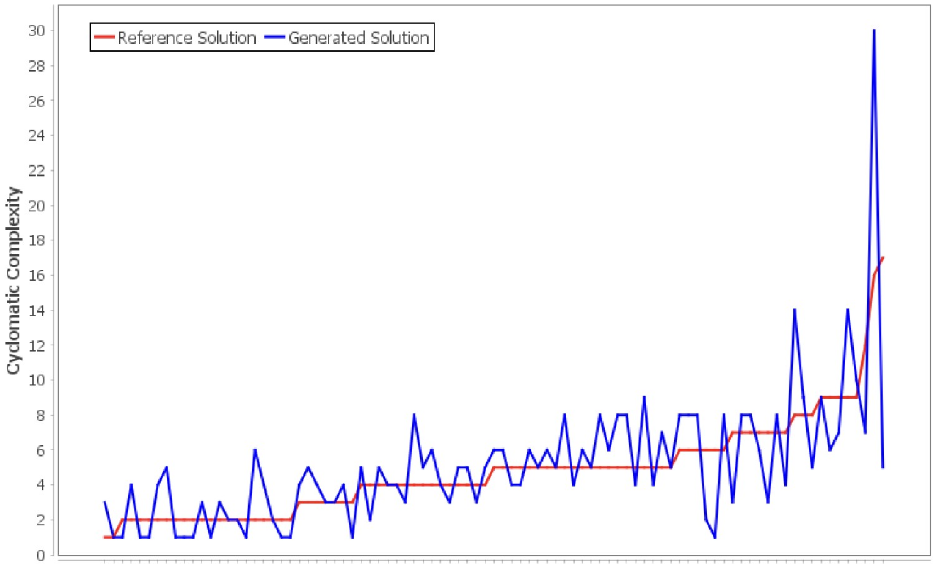}\\
\scriptsize{(a) Comparison on Cyclomatic Complexity}\\
\includegraphics[width=8.5cm, height=3cm]{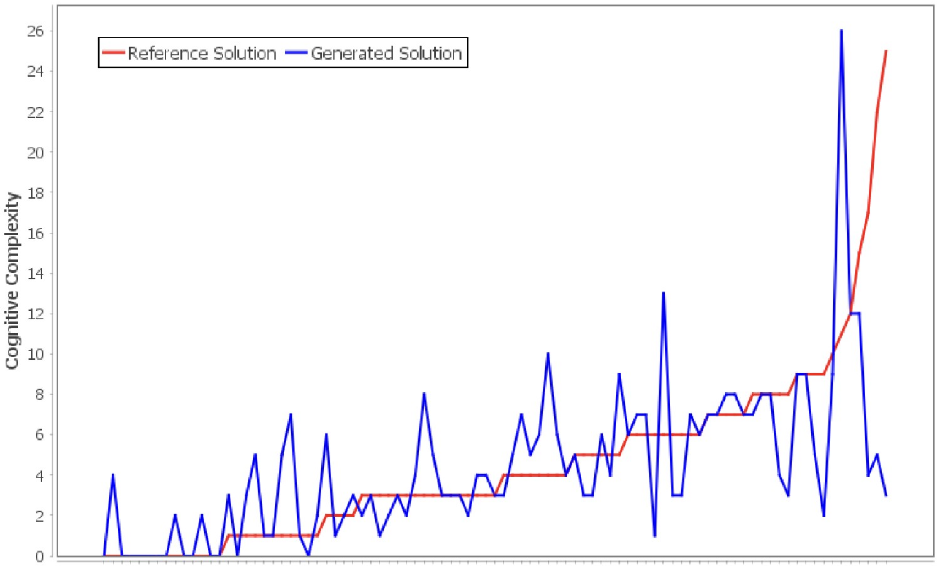}\\
\scriptsize{(b) Comparison on Cognitive Complexity}\\
%\includegraphics[width=8.5cm, height=3cm]{figs/figLOC.pdf}\\
%\scriptsize{(c) Comparison on Lines of Code}\\
\captionsetup{justification=centering, position=above}
\caption{Complexity of Reference Vs Generated Solutions}
\label{fig28_30}
\end{figure}

To analyse the quality of the generated code, we compare the complexity of the generated solutions with that of the reference solutions. We construct a random subset of the benchmark, and submit the tasks of the dataset to ChatGPT. The results from ChatGPT are analysed by invoking the test morphism \emph{analyseComplexity}. Figure \ref{fig28_30} shows the distributions of cyclomatic and cognitive complexities of the reference solutions and the generated codes, where red lines are the complexities of reference solutions while the blue lines are the complexities of generated codes. 

\begin{table}[b]
\centering
\caption{Complexity of Generated vs Reference Solutions}\label{tab:complexityAll}
\begin{scriptsize}
\begin{tabular}{|c|c|c|c|c|c|c|}
\hline
\textbf{Complexity}&\textbf{AvgRS}&\textbf{AvgGS}&\%\textbf{Above}&\%\textbf{Equal}&\%\textbf{Below}&\textbf{Avg}$\delta$\\
\hline
Cyclomatic &5.36&5.52&48.45&18.56&32.99&2.31\\
\hline
Cognitive&5.21&5.01&41.24&25.77&32.99&2.62\\
\hline
LOC&33.52&24.16&33.03&7.24&58.82&14.47\\
\hline
CLOC&39.18&24.84&24.89&4.52&69.68&18.29\\
\hline
CL&5.72&0.84&14.93&27.15&57.014&5.59\\
\hline
\end{tabular} 
\end{scriptsize}
\end{table}

\begin{table*}[t]
\caption{Complexities of Generated and Reference Solutions for Correct and Incorrect Tasks}\label{tab:ComplexityCorrectness}
\centering
\begin{tabular}{|c|c|c|c|c|c|c|c|c|c|c|c|c|}
\hline 
 & \multicolumn{6}{c|}{\textbf{Correct Subset}} & \multicolumn{6}{c|}{\textbf{Incorrect Subset}} \\ 
\hline 
\textbf{Complexity} &\textbf{AvgRS} &\textbf{AvgGS} &\textbf{\%Above} &\textbf{\%Equal} &\textbf{\%Below} &\textbf{Avg$\delta$} &\textbf{AvgRS} &\textbf{AvgGS} &\textbf{\%Above} &\textbf{\%Equal} &\textbf{\%Below} &\textbf{Avg$\delta$}\\ 
\hline 
Cyclomatic &5.25  &5.69  &60.66  &13.11  &26.23  &1.92  &6.00  &4.81  &27.78  &27.78  &44.44  &2.97\\ 
\hline 
Cognitive    &5.26  &5.36  &39.34  &29.51  &31.15 &2.23  &5.14  &4.69  &36.11  &19.44  &44.44   &3.67\\ 
\hline 
LOC            &31.87 &23.97 &34.23 &9.01  &54.95  &12.45 &35.15 &24.34 &31.82 &5.45 &62.73   &16.48\\ 
\hline 
CLOC         &40.19 &24.52 &15.32 &3.60  &79.28  &17.16  &38.17 &25.15 &34.55 &5.45 &60.00    &19.41\\ 
\hline 
CL              &8.43   &0.74   &3.60   &6.31   &88.29  &7.80   &3.04   &0.93   &26.36  &25.45  &48.18 &3.40\\ 
\hline 
\end{tabular} 
\end{table*}

Statistical analysis of the complexity data is given in Table \ref{tab:complexityAll}, where columns AvgRS and AvgGS are the average complexities of the reference and generated solutions, respectively. The columns \%Above, \%Equal and \%Below are the percentages of the tasks that the generated solutions have a higher, equal and lower complexity than the reference solutions, respectively. The column Avg$\delta$ gives the average absolute differences between the complexities of the generated and reference solutions. The rows of Cyclomatic and Cognitive are the data of cyclomatic and cognitive complexities. The row LOC gives the data about the numbers of lines of the code after comments are removed, CLOC is about the number of lines with comments, and CL is about the number of lines that contain comments. 

From the data given in Table \ref{tab:complexityAll}, the generated solutions are often more complex than the reference solutions. The average cyclomatic complexity of the generated solutions was 5.52, while the average of reference solutions was only 5.36.  On 48.45\% of the tasks the generated solutions have higher cyclomatic complexity than reference solutions, while only 32.99\% tasks have lower cyclomatic complexity, and 18.56\% tasks were of equal complexity. For cognitive complexity, the generated solutions have a slightly lower average cognitive complexity than that of the reference solutions. However, there are more cases that generated solutions have a higher cognitive complexity than reference solutions. 

Our experiment data also shows that the generated codes are much shorter than the reference solutions in terms of the number of lines of code. On average the generated code is more than 14 lines shorter. This contradicts the observation made by Miah and Zhu on the R program code generated by ChatGPT \cite{c15}.  Moreover, the generated code contains little comments. On average there is only less than 1 line in the generated code. 

To further investigate the complexity of ChatGPT generated code, we split the test dataset into two subsets according to the functional correctness: one contains the tasks that the generated code passes all test cases, and the other for those that the generated code fails on tests. The statistical data is shown in Table \ref{tab:ComplexityCorrectness}. 

The data in Table \ref{tab:ComplexityCorrectness} show that the correctly generated solutions are likely to be more complex than reference solutions in both cyclomatic and cognitive complexity metrics, and longer as well. However, for incorrectly generated codes, it is more likely to be less complex than the reference solution and more likely to be shorter. 

\subsection{Discussion}

From the experiments, we make the following observations. 

First, due to the metadata associated with the task in the ScenEval benchmark and the support of datamorphic testing tool Morphy, scenario-based testing and evaluation can be conducted efficiently and effectively. Benchmarks with metadata and a carefully developed test system with test morphisms for dataset filtering, test result analysis and test data distribution analysis form a powerful scenario-based test and evaluation environment, in which experiments can be conducted efficiently and effectively. 

Second, using scenario-based evaluation, one can gain insight into an LLM model effectively. For example, we can identify the task topics on which ChatGPT performed poorly. These topics are the areas that ChatGPT should improve. It has been observed already that when the complexity of the tasks increases, the LLM's performance decreases. However, existing works are based on an informal judgement of the difficulty of tasks \cite{c15}. In our experiments, the complexity of tasks are measured by cyclomatic complexity and performances are evaluated on subsets of different complexities. It is further confirmed that the decrease in performance on complexity is not a coincidence because the complexity of tasks in certain topics are more complicated than other topics. 

Finally, the Morphy testing tool makes the test system easy to manage and operate and flexible to extend and evolve. Various software engineering tools can be easily integrated together via implementation of invocations of existing tools and code to extract data saved by such tools. Our test system has integrated PMD and EvoSuite tools. 

\section{Conclusion}\label{secVI}

We proposed a new approach to structure benchmark datasets with metadata to represent the usage scenarios of each element of the benchmark and to develop a test system for using metadata to support scenario-based testing and evaluation of LLMs. We have demonstrated how metadata makes it possible to formulate scenarios efficiently and how scenarios contribute to a thorough analysis of LLM performance.

For future work, we are further developing the test system with more test morphisms for analysis of the quality of program code. The work reported in this paper has only considered correctness and complexity. Many other aspects of code qualities can and should be evaluated, and this is already in progress. 

We are also working on the testing and evaluation of many other LLMs for code generation. This can be implemented fairly easily by writing a test morphism for  executing  each LLM to invoke the model with queries extracted from the description of the task and to receive the responses from the LLM. Comparisons between many LLMs could be time consuming and labour intensive when the process involves manual processing of data. We are working on how to automate the whole process. 

\addtolength{\textheight}{-10.0cm}   % This command serves to balance the column lengths

%%%%%%%%%%%%%%%%%%%%%%%%%%%%%%%%%%%%%%%%%%%%%%%%%%%%%%%%%%%%%%%%%%%%%%%%%%%%%%%%

\end{document}